\documentclass[11pt]{article}
\usepackage{setspace}
\usepackage{amssymb,verbatim,epsfig}
\usepackage{amsmath,setspace}
\usepackage[comma,sort&compress,super]{natbib}
\usepackage{multicol}
\usepackage[usenames,dvipsnames]{color}
\usepackage{url}
\usepackage{amssymb, amsmath, verbatim, epsfig}
\usepackage{setspace}
\usepackage{lineno}
\usepackage{xcolor}

\newcommand{\fig}[1]{{\bf Figure \ref{#1}}}
\newcommand{\tab}[1]{Table \ref{#1}}

\usepackage{mathtools}

\spacing{1}    
\usepackage[hang,small,bf]{caption}

\title{\bf 
\vspace{-1in}
Conservation Tools: The Next Generation of Engineering--Biology Collaborations}

\author{Andrew K. Schulz$^{1,2,+,*}$, Cassie Shriver$^{3,+}$, Suzanne Stathatos$^{4,+}$, \\ Benjamin Seleb$^{3}$, Emily Weigel$^{3}$, Young Hui Chang$^{3}$,\\ M. Saad Bhamla$^{5}$, David Hu$^{1,3}$, Joseph R. Mendelson III$^{3,6}$ \\
\text{\small{Schools of Mechanical Engineering$^1$, Biological Sciences$^3$, and Chemical and Biomolecular Engineering$^5$}}\\
\text{\small{Georgia Institute of Technology, Atlanta, GA 30332, USA}} \\
\text{\small{Max Planck Institute for Intelligent Systems$^2$, Stuttgart, Germany}} \\
\text{\small{School of Computing and Mathematical Sciences$^4$}} \\
\text{\small{California Institute of Technology, Pasadena, CA 91125, USA}}\\
\text{\small{Zoo Atlanta$^6$, Atlanta, GA 30315, USA}}
\text{\small{}}}

\begin{document}
 \maketitle

\noindent \text{\small{+ indicates co-first author}} \\
\noindent{\bf Corresponding author:} \\
Andrew Schulz \\
Heisenbergstraße 3, Stuttgart, Germany 70569\\
(405)780-0542 \\
andrew.schulz1994@gmail.com\\

\section*{Abstract}
The recent increase in public and academic interest in preserving biodiversity has led to the growth of the field of conservation technology. This field involves designing and constructing tools that utilize technology to aid in the conservation of wildlife. In this article, we will use case studies to demonstrate the importance of designing conservation tools with human-wildlife interaction in mind and provide a framework for creating successful tools. These case studies include a range of complexities, from simple cat collars to machine learning and game theory methodologies. Our goal is to introduce and inform current and future researchers in the field of conservation technology and provide references for educating the next generation of conservation technologists. Conservation technology not only has the potential to benefit biodiversity but also has broader impacts on fields such as sustainability and environmental protection. By using innovative technologies to address conservation challenges, we can find more effective and efficient solutions to protect and preserve our planet's resources. 

\section*{Background and Motivation}
The term "conservation technology" was first proposed by Berger-Tal in 2018\cite{bergertal_conservation_2018} to broadly describe the use of technology to manage and conserve wildlife. While a commonly referenced example is unmanned aerial vehicles (UAVs, also known as drones), there are many other conservation technologies, including camera traps, mobile applications, spatial mapping, and environmental DNA. Much of the existing technology uses modern hardware and software design processes to improve upon ongoing conservation efforts and initiate previously under-addressed efforts\cite{bergertal_conservation_2018}. Some of the major goals of conservation technology are to iterate more quickly to improve outdated equipment, increase accessibility to tools, and use modern technology to address conservation problems in entirely new ways. Conservation technology is being developed for animals in both natural environments and captive settings (e.g. foxes in urban settings and elephants in zoos, respectively) \cite{pacheco_how_2018} and applications for this technology may include conservation challenges and monitoring needs for animals, plants, habitats, geological phenomena such as volcanoes, climate, and atmosphere, and more. 

Why has there been a revolution in implementing these new and old tools in the conservation sector in the last few years? The recent broader recognition of the significant threats to biodiversity has driven demand for conservation technology. Since 1970, wildlife populations have plunged by 69\%\cite{wwf_report_2022}. With advancing technological revolutions over the past millennium, many scientists, engineers, and other conservation stakeholders see conservation tools as valuable methods to address serious ongoing conservation challenges.

Historically, the field of conservation technology has taken an opportunistic approach wherein stakeholders invest in developing technology for a specific, non-conservation need that is then applied to wildlife management\cite{bergertal_conservation_2018}, a prime example being the development of drones by the military. While opportunistic technologies certainly aid in wildlife management efforts, they also tend to be expensive, less accessible to the conservation community, and rarely the most idealized solution for specific wildlife management issues. Given the alarming rates of biodiversity loss amidst the current mass extinction\cite{ceballos_accelerated_nodate}, there has been an increasing push towards purpose-driven technology designed in consultation with members of the conservation community\cite{bergertal_conservation_2018}. Critically, new technology is not considered to represent proper conservation technology until its genuine usefulness and success for managing and conserving wildlife has been demonstrated.

Producing purpose-built technology requires a variety of skills that are typically beyond the scope of a single person, so establishing successful interdisciplinary collaborations is crucial. To properly synthesize these perspectives, conservation technology must establish the necessary bridges between the conservation community, technologists, and policymakers \cite{sintov_fostering_2019,bergertal_conservation_2018}. However, these interdisciplinary collaborations are dependent on effective communication across domains, which can be difficult given differences in objectives and goals. While technology development and outcomes are often derived from an engineering design mindset, biological conservation is more hypothesis-driven and grounded in the scientific method. 

Despite grants, training, and other efforts towards these collaborations, the conservation community has encountered many solutions claiming to be universally minded but lacking in necessary interdisciplinary knowledge and partners in respective fields. Inadequate communication between fields initially led the wildlife community to be distrustful of new technologies because many engineering solutions were poorly applied to serve the needs of conservationists, especially in natural outdoor situations. However, the pressing nature of the sixth mass extinction and climate change has made the necessity of interdisciplinary solutions evident and worth pursuing despite communication difficulties. And with this expansion of interdisciplinary collaborations comes the realization that novel contributions to conservation technology can be designed in a variety of ways.

The term "conservation technology" has received much criticism from the conservation community for the implication of requiring advanced technologies. This is misleading when plenty of modern innovations involve using simple non-hardware/software devices to serve novel conservation problems. Conservation technology can be as complex as machine learning used to identify and track species or as simple as chili pepper fences used to deter African Elephants from damaging farmland\cite{tuia_perspectives_2022,changa_scaling-up_2016}. It is for this reason that we will utilize the term \textbf{conservation tools} (CT) instead of conservation technology for the remainder of this manuscript. A tool is broadly defined as a device to carry out a particular function, and we believe using the term "conservation tools" better encompasses the diversity of the field. Rephrasing also intentionally includes indigenous solutions utilized in traditional conservation practices around the globe, which may not be accurately described in the scope of conservation technology.

In this manuscript, we will describe a diverse array of case studies of \textbf{conservation tools} that have been implemented globally. We also emphasize how the importance of the project to work alongside the communities impacted with the communities as the stakeholder to minimize the traditional practices of parachute science while maximizing community partnerships, access, and conservation impact. Parachute science, where scientists "parachute" into places for purposes of research or conservation but leave without a trace of co-authorship to community members who made the work possible, is an unfortunately common phenomenon\cite{schulz_guide_2022}. We describe tools that are heavily focused on hardware and heavily focused on software but can still be simple to use, build, and adapt to additional conservation needs. This manuscript is meant to be a starting guide to introduce the field of creating conservation tools to those without experience. We hope that the glossary of terms also allows current practitioners of CT to understand the diversity of this field better. This manuscript hopes to allow the reader to understand that biodiversity is essential not just in the wild but also in the technologies utilized to help the conservation tools be as effective as possible. 


\section*{Conservation Tools Vocabulary}
As the conservation technology field has grown, it has adopted many terms from other fields to describe conservation tools accurately. Unfortunately, many of these technical terms are domain-specific and can alienate stakeholders. We list and define many of the terms commonly used to describe conservation tools and provide corresponding publications with more details on these specific terms (\tab{tab:Table2_1}). We elaborate on each term in the case studies and introduction. Throughout this paper, we present these terms in bold font to indicate where readers can refer to this table for additional details. 
 
\section*{Discussion}

We propose to shift the phrase "conservation technology" to conservation tools because the traditional process of creating conservation technology relies on what is known as opportunistic technology\cite{bergertal_conservation_2018}. Purpose-built technology is common in the hardware and software industry, considered collectively under the  term \textbf{Human-Centered Design} (HCD). HCD operates by using a design mindset that focuses on the context of the use of the idea. A common example is the difference between checkout interfaces in different environments. Purchasing a beverage at a bar versus at a supermarket are similar situations, but the context of use for exchanging money for goods is completely different. In each scenario, there are a set number of individuals, \textit{m}, and a set number of items each individual has, \textit{n}. For the bar, there is a very large \textit{m} with a very small \textit{n}, but the opposite is true of the supermarket. Thus the design of the technology and processes that enable the purchasing of goods are in very different contexts of use. We will apply the same logic to conservation tools in utilizing a \textbf{Human--Wildlife-Centered Design} (HWCD) approach. 

A HWCD approach for conservation tools requires consideration of not just the human interaction with the device but also the interaction between humans and wildlife. While "human-wildlife conflict" \cite{soulsbury_humanwildlife_2015} is used to describe interactions in urban, farming, or wild settings that can cause large amounts of harm to human interests\cite{jessen_contributions_2022}, "Human-Wildlife interaction" is broadly used to describe both positive and negative interactions. HWCD is not a new concept, as it has been implemented for millennia by indigenous peoples that live, interact, and move with the land. For designers from non-indigenous backgrounds, it is essential to understand that you will never be able to achieve true indigenous design unless the primary designers of a technology solution are from the native indigenous lands where the solutions will be implemented\cite{the_ecological_society_of_america_ecological_nodate}. To ignore indigenous or other community-derived knowledge is to create a solution with only partial expertise or knowledge of the problem; for this reason, we implore readers to understand that the effectiveness of your tool relies on the active collaboration of the community, scientists, and engineers. As we go forward in this manuscript, it is paramount for authors to understand that the best tools are created by indigenous researchers, scientists, and engineers working collaboratively as they are the most knowledgeable folks in the world about the conservation challenges non-indigenous members outside the community have imposed.

\begin{figure}
	\includegraphics[width=1\textwidth,page=1,trim=0.0in 0in 0in 0in,clip]{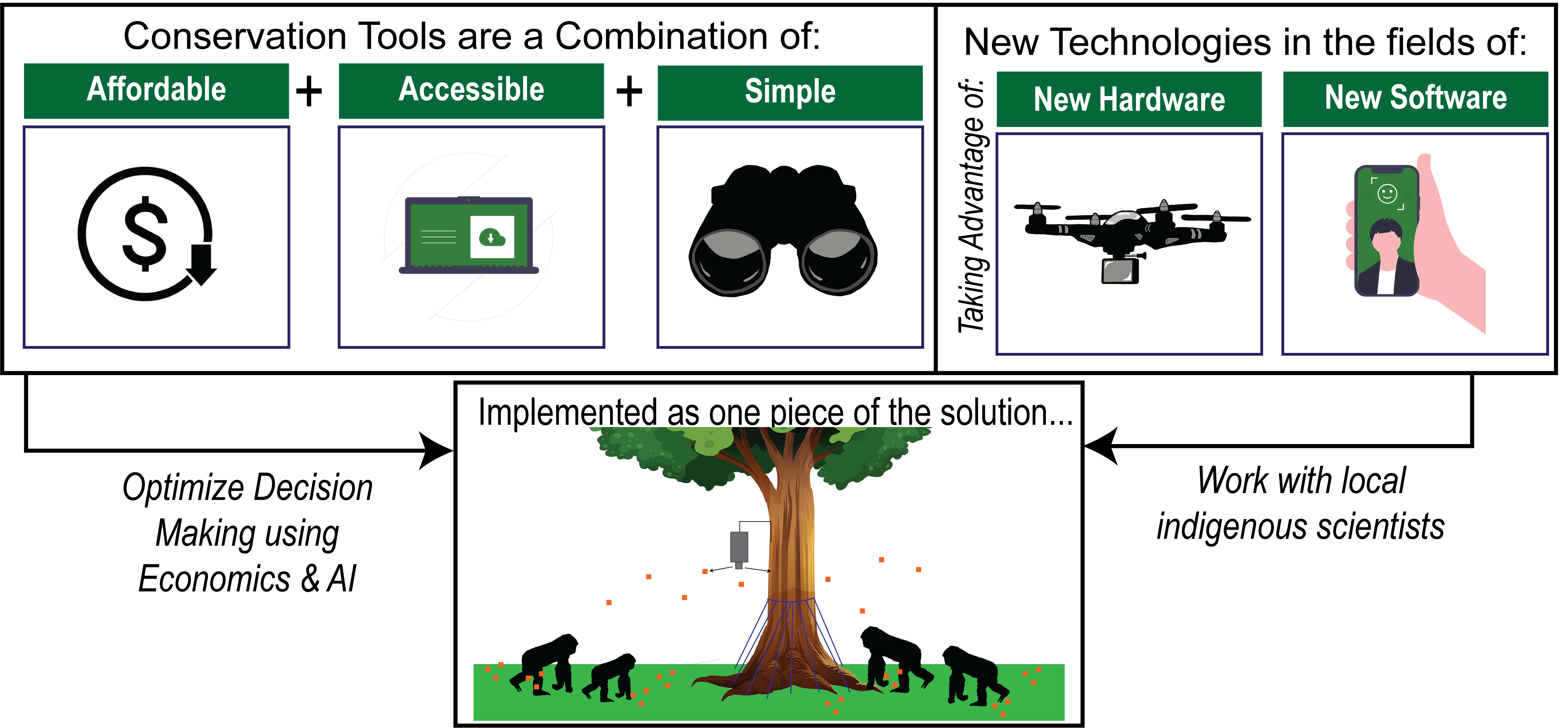}
	\centering
	\caption{Visual Abstract displaying the Conservation Tool framework discussed in this piece. Silhouettes created by Gabriela Palomo-Munoz and Undraw.co.\\}
	\label{CTperspectivesFig}
\end{figure}

We will now discuss five case studies to introduce the core principles which we highlight in \fig{CTperspectivesFig}, which we believe conservation technology solutions should be implemented. The themes are listed at the beginning of each section. 
\begin{itemize}
\item What – what is its use?
\item How – how is it used?
\item Where – what are the use cases/how it helps
\item Why – future directions/open questions/etc. 
\end{itemize}

The following case studies cover specific conservation tools that are created, drawing from both new and old technologies. Each of these solutions utilizes some measure of HWCD., although some utilize frugal materials and are  simple, while others take advantage of advanced hardware and software that have become more accessible in recent years. The themes of these case studies relate to what the surveyed community of conservationists believe are the most important tools for assisting in advancing conservation from the most recent state of conservation technology report\cite{speaker_global_2022}. We proceed with discussing a case study on accessibility in technology. 

\subsection*{Case Study 1: AudioMoth}
\noindent \textbf{Principle: Solutions should be open source, and accessible in cost and function}\\
Open-source software often describes the ability to access the code and customize and edit the code how we see fit. Undergraduate biologists are often taught R, an open-source programming language geared toward statistical modeling; comparatively, engineers and computer scientists frequently learn Python, an open-source all-purpose programming language. Regardless of the coding language used, open-source code can then be appreciated by the collaboration community where the prior knowledge only differs by which open-source software is used in their curriculum. Workshops\cite{cv4ecology} and online forums\cite{wildlabs} have begun to bridge the gap between the two groups; for example, in the CV4Ecology workshop, engineers teach conservation biologists at the graduate level and post-doctoral levels specific tools in Python. One example of how open-source software and hardware are used for conservation is AudioMoth.

\textbf{What is its use?} Effective wildlife management decisions require abundant data on the organisms. Acoustic monitoring has become one of the more ubiquitous methods of recording information in field situations where sound is relevant\cite{obrist_rapid_2010,blumstein_acoustic_2011}. While early practices required individuals to actively note the sounds they heard, passive monitoring devices can be deployed into areas of interest to record information for animals located within a certain proximity\cite{obrist_rapid_2010,blumstein_acoustic_2011, sugai_terrestrial_2019}. 

\begin{figure}
	\includegraphics[width=1\textwidth,page=1,trim=0.0in 0in 0in 0in,clip]{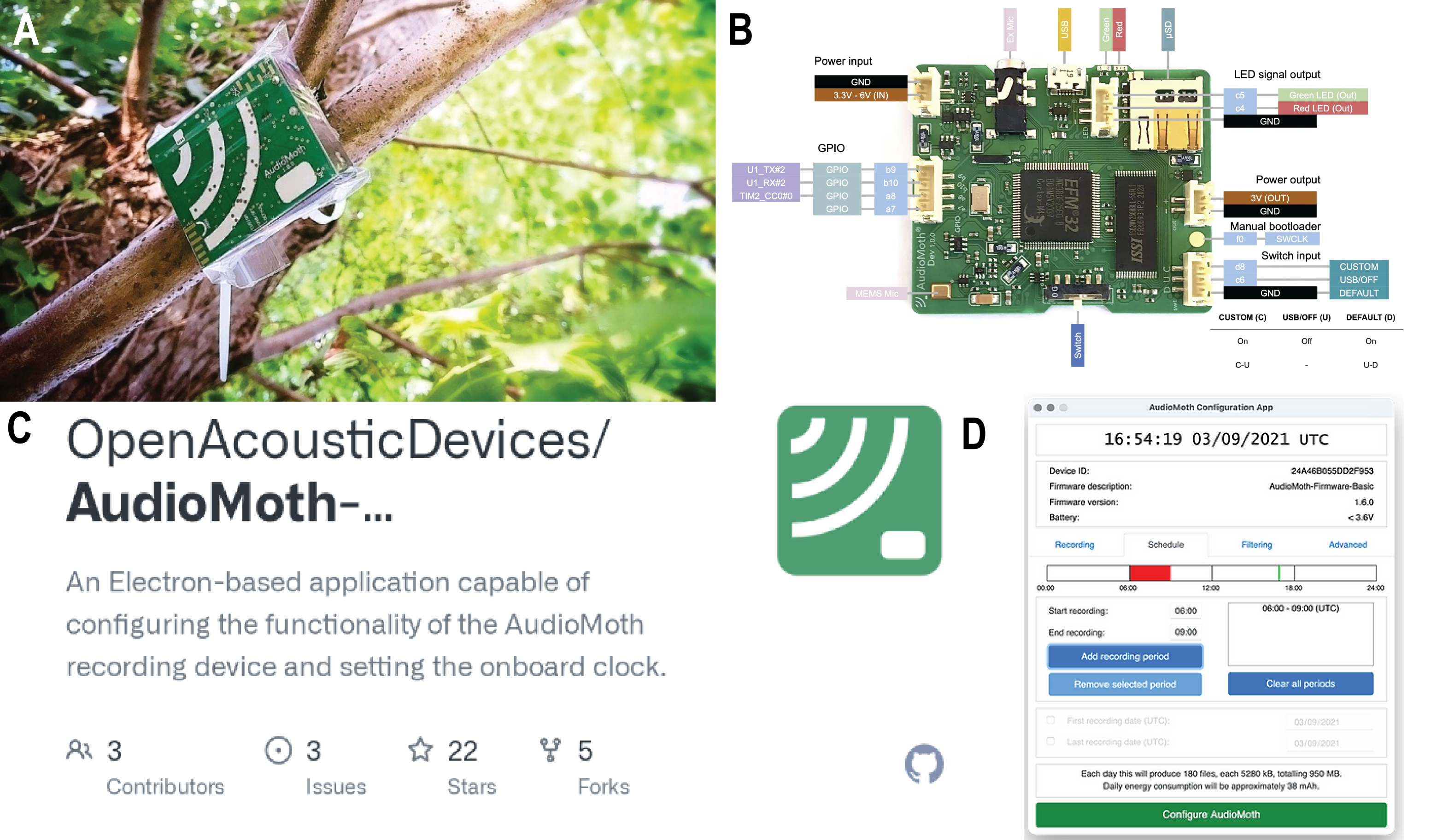}
	\centering
	\caption{A) Audiomoth, B) Open source printed circuit board that audiomoth includes on website, C) Open source code for controlling and interpreting data from the audiomoth via Github, D) Online and app-based user-interface for audiomoth users. Images were taken from AudiomMoth website with permission.\\}
	\label{AudioMoth}
\end{figure}

\textbf{How is it used?} The AudioMoth costs ten times less than commercial products, is energy efficient, records both human-audible sounds and ultrasonic frequencies, is the size of a credit card, and was created by two PhD students with the intention of increasing scientific accessibility(\fig{AudioMoth}A)\cite{hill_audiomoth_2019} . Furthermore, this device is open-source, meaning the code and operating features are made public for distribution and modifications for individual projects\cite{hill_audiomoth_2018}. Immediately popular, it has been used to monitor animal populations\cite{revilla-martin_monitoring_2021}, track migrations\cite{roark_monitoring_2021}, identify poaching activity \cite{hill_audiomoth_2018}, detect sounds underwater\cite{lamont_hydromoth_2022}, and even discover new species\cite{hill_audiomoth_2018}. 

\textbf{What is a use case?} The AudioMoth exemplifies how designing technology that is open-source and accessible can dramatically increase scientific participation, with substantial implications for informing future wildlife management policies and practices. The term open source solution can mean several different things when looking at a device such as AudioMoth and we will discuss these in a set of categories: open-source hardware, open-source software, and open-source code.

These devices can substantially increase monitoring coverage both in terms of land area and recording time\cite{obrist_rapid_2010,blumstein_acoustic_2011, sugai_terrestrial_2019}. While the multi-functional device has applications throughout the biological world it has seen greater use recently with the release of the United Nations's sustainable development goals (SDGs), specifically investigating Life below Water (SDG 14) and Life on Land (SDG 15). However, initial productions were far too expensive and complex for mass implementation in the scientific community until the creation of the AudioMoth. 

A core feature of open-source hardware is that the project can be built of mechanical parts that are all easy to acquire. This can mean a variety of things from easy to purchase or easy to create using different advanced manufacturing techniques such as 3D printing or laser cutting. specifically 3D printing, sometimes referred to as additive manufacturing, allows for specific parts of a hardware model to be built at low expense\cite{pearce_building_2012}. Truly open-sourced hardware will not just tell you the techniques utilized but will also include the exact parts, files, models, etc. required for this. A new journal publication type is leveraging the future of open-source hardware through publications. These journals currently include the \textit{Journal of Open Hardware, The Journal of Open Engineering, and HardwareX}. These journals require all submissions to include complete information for all hardware and software included in the device\cite{pearce_economic_2020}. It should, however, be mentioned that some of these publishers, including Elsevier, are for-profit publishers. 

\textbf{What is the Potential?}
Open-sourced hardware does not just mean the mechanical devices such as nuts and bolts, but also the electrical devices, such as the circuit board and circuitry diagram (\fig{AudioMoth}B). By providing these schematics, users can build the entire AudioMoth system using the specifications sheets provided in their publication in HardwareX. There can be large gaps in the idea of open source between hardware and software. Many devices allow you full access to the sourcing and schematics of the hardware but require you to purchase specific software from the organization. 

 Two of the commonly applied to be used in this context are front-end interface and back-end interface. The front-end interface is what the primary user sees on a screen, or the user-interface (UI)\cite{smith_professional_2012} . The back-end interface is internal equipment that is actually doing much of the coding work\cite{smith_professional_2012}. Many of the devices will not feature a customizable front end because it will directly reflect a customizable back end known as the application programming interface or API. An API is a primary way of allowing open-source code not only to be accessed but also updated and the outputs to be changed. This is necessary for researchers because acquiring an API lets the researcher customize the data collection, for example extracting location information, or measurements of temperature or velocity all of which allow the open-source tool to be customizable for both inputs and outputs. The manufacturer of AudioMoth, OpenAcoustics (https://www.openacousticdevices.info/audiomoth), provides the researcher not only an API but also all of the operational code, in a user manual for each device thus permitting customization of any device aspect (\fig{AudioMoth}C-D). Finally, its ease of use for the end user is a crucial design component. In designing this device the engineering developers have a feedback mechanism for those in the field to help continuously improve the use of this. The engineers and scientists developed this tool to be used by indigenous researchers and people in which they could place these devices wherever they see fit to start receiving data. Those that have occupied indigenous lands are the most knowledgeable about where the placement of these devices would be most effective. 

In designing a tool with HWCD in mind, it is vital to think of the context of use. The AudioMoth is intended to be used by ecologists attempting to get bio-acoustic data from their open-source sensor. Biologists differ significantly from computer scientists and engineering researchers;  biology is a hypothesis-based field, whereas engineering is design based. The AudioMoth is designed by and for biologists to be deployed quickly and repaired easily in various applications and environments. AudioMoth utilizes advanced technology in both software and hardware while utilizing context-of-use to consider how it will be used.

\subsection*{Case Study 2: Environmental DNA (eDNA)}
\noindent \textbf{Principle: Solutions should take advantage of increasing hardware technologies}\\
DNA is a well-established scientific tool for an ever-expanding scope of biological studies and beyond\cite{fair_expanding_2021}. An enormous challenge in the use of DNA for purposes of conservation is that traditional methods of DNA collection require biological samples such as urine, hair, skin, or other tissue\cite{dairawan_evolution_2020}. Traditional biological methods have historically required the restraint, capture, or rapid collection of fresh DNA samples that can be either logistically infeasible or actively at-odds with observing organisms in the wild. Focal organisms in many conservation programs often are extremely rare or secretive, and it may not be possible or logistically feasible to get samples from them. Thomsen and Willerslev (2015) reviewed the use of eDNA as an emerging tool in conservation\cite{thomsen_environmental_2015}. One of the primary challenges they highlighted in conservation is the trade-off between the invasiveness of studies and data collection. Applications of eDNA are reducing the needs for invasive studies and enabling locating and monitoring of creatures too rare or secretive for traditional survey methods. 

\textbf{What is its use?} Environmental DNA allows for the analysis of diets, geographical ranges, population sizes, demographics, and genetics, as well as the assessment of the presence/absence of species at sites. These can be quantified using environmental samples such as feces left behind and analyzed for different genetic information. The field of eDNA benefits the conservation space as being a holistically non-invasive method of DNA extraction, making it very repeatable. The techniques for the collection of eDNA still require biological sample collection, but the samples can be in much lower quantity and make use of vacuums to process the needed concentrations. 

\textbf{How is it used?} This is a previously developed tool, more recently used by wildlife conservationists, which allows using DNA samples found in the environment for understanding endangered species in their natural environment utilizing only DNA samples. The novelty in this tool is its ability to not only detect DNA information about animals, but it is applicable across a variety of environments including land, sea, and in polar ice samples\cite{thomsen_environmental_2015}. As a tool, eDNA is useful in a variety of conservation and ecological fields, but this solution has had a significant history of colonial-style parachute science\cite{von_der_heyden_environmental_nodate}. It is important to note that while solutions and technology like this can be leveraged, they must be thought of in the HWCD framework. Working with local and indigenous communities as eDNA is not the sole solution and additionally on the ground conservation work is necessary for the long-term conservation of wildlife\cite{lacoursiere-roussel_environmental_2021}. 

\textbf{What is a use case?:} One of the first documented uses of eDNA was in 1992 when Amos utilized shed skin from cetacean mammals species to inform a population analysis\cite{amos_restrictable_1992}. Although not considered as conservation technology at the time, this was one of the first applications of non-invasive eDNA for biological conservation and population assessment. Now eDNA is utilized to monitor not just populations but to catalog local bio-diversity of fishes\cite{shen_edna_2022}, manage reptile populations\cite{nordstrom_review_2022}, and forest conservation\cite{lock_harmonizing_2022} using the interface between remote sensing and environmental DNA.

\textbf{What is the potential?} 
This solution appears to be a universal (or cure-all) tool. Universally designed solutions typically will only work for specific use cases. The non-universality in eDNA is that it leverages the well-established scientific tool of DNA for new and innovative applications for purposes of conservation data and management. Despite its broad range of potential applications, eDNA nevertheless is not a universal solution for all situations, especially because the methodology is complex in terms of sampling and acquisition of data. As in a wet lab technique it is prone to the same human errors as other lab based risks including contamination, biased results and interpretations, or even as simple as not having adequate reference databases for identifying DNA sequences for all regions or applications.

These pitfalls do not discount eDNA as an example of utilizing new advances in hardware and scientific progress to advance conservation practices. Tools such as this continually are being improved and innovated. This field is expanding in the past years with increasing establishments of DNA Barcodes that permit identification of species using online DNA databases\cite{gostel_expanding_2022}. 

\subsection*{Case Study 3: Computer Vision}
\noindent \textbf{Principle: Solutions should take advantage of increasing software technologies}\\
{\bf Machine learning} is the science and art of developing computer algorithms to learn automatically from data and experience\cite{cs155_lec1}. \textbf{Computer vision} is a sub-field of machine learning, in which computers and systems are trained to extract meaningful information (aka "see") from images, videos, and other inputs. Computer vision lets computers understand visual inputs\cite{cv_book}.

\textbf{What is its use?}  While humans have been ``trained" during their lifetime to identify objects, understand their depth, and see their interactions, computer models require thousands of images to teach machines to ``see'' new scenarios. Computer vision has expanded in recent years, too, from only being able to work on super-computers to now working on edge devices like cell phones and laptops in the wild\cite{reed_reinventing_2022, gvh_inat_2018,  gvh_merlin}. With hardware advances and algorithms designed for lower-resource devices, computer vision has become less expensive and more accessible to many organizations that wish to use it. Users of computer vision applications today include, but are not limited to, (1) iPhone users to unlock their phones with their face, (2) drivers of self-driving cars, and (3) traffic enforcers who use red-light traffic cameras.

\begin{figure}[h]
    \centering
    \includegraphics[width=0.7\textwidth,page=1,trim=0.0in 0.0in 0in 0.0in,clip]{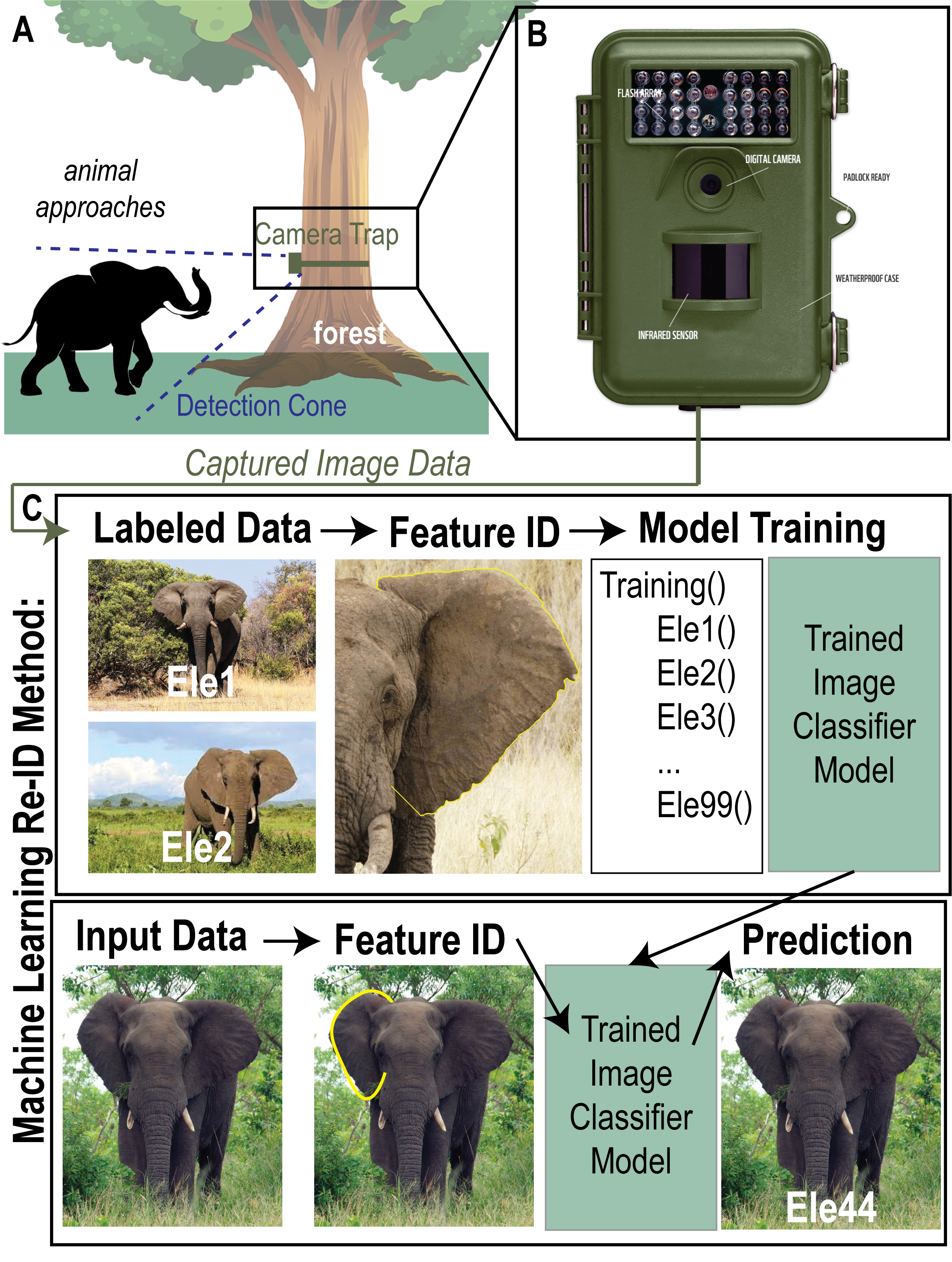}
	\centering
    \caption{\textbf{Basic camera trap setup}. A) A camouflaged camera trap is often placed on a tree or a pole. B) Camera trap model. It is equipped with a motion-triggering sensor, a digital camera, and a memory card. When an animal passes in the region of interest, the camera captures photos/video at a specified frame rate of the animal. C)  The figure was made using a dataset from LilaBC \cite{lilabc_leopards} and images from Flickr.\label{CameraTrap}}
\end{figure}

\textbf{How is it used?} Conservationists use camera traps to capture images of wildlife. Computer vision techniques are applied to these camera trap images to help scientists detect, track, classify, and re-identify (recognize) individual animals, among other things \cite{beery2020iwildcam, beery2018recognition}. A typical camera trap apparatus is shown in \fig{CameraTrap}A. A camera is placed in a region of interest. It passively collects information about what goes through that region. Camera traps collect data over a specified period, either writing to an external hard drive or pushing data to a cloud-hosted framework. Camera traps often include infrared and/or motions sensors that can identify warm-bodied or moving objects. When an animal triggers the sensor, the camera records (writes images to memory), as shown in \fig{CameraTrap}B. Afterward, the data is fed into a machine learning model to learn and recognize patterns in the data, \fig{CameraTrap}C. 

Traditionally, computer vision has used classical supervised machine learning algorithms (algorithms that need human-labeled data). These algorithms let the model understand identifying characteristics of the animals within the regions of interest (for example, color histograms, texture differences, locomotor gait, etc. \fig{CameraTrap}C) \cite{tuia_perspectives_2022}. From those characteristics, the model can learn to detect and classify wildlife species in the images. Two key use cases of this include classifying animal species ({\bf classification}) and recognizing and identifying individual animals ({\bf re-identification or re-id}). In these tasks, extraction of the foreground of the image is an important pre-processing step to focus the model on the animal of interest. For example, the first step in classifying urban wildlife is often to crop the image to focus on the animal and not to focus on cars, leaves, trees, etc .\cite{beery2019efficient}. The model could then take these cropped images and classify them as different species types (i.e. squirrels, dogs, coyotes, etc.).  Alternatively, \cite{beery2019efficient} could be used to identify which photos to ignore. For example, several urban wildlife monitoring projects use it to crop humans out of images and ignore empty images. \cite{urbanwildlife}. 

\begin{figure}
    \centering
    \includegraphics[width=1\textwidth,page=1,trim=0.0in 1in 0in 0.3in,clip]{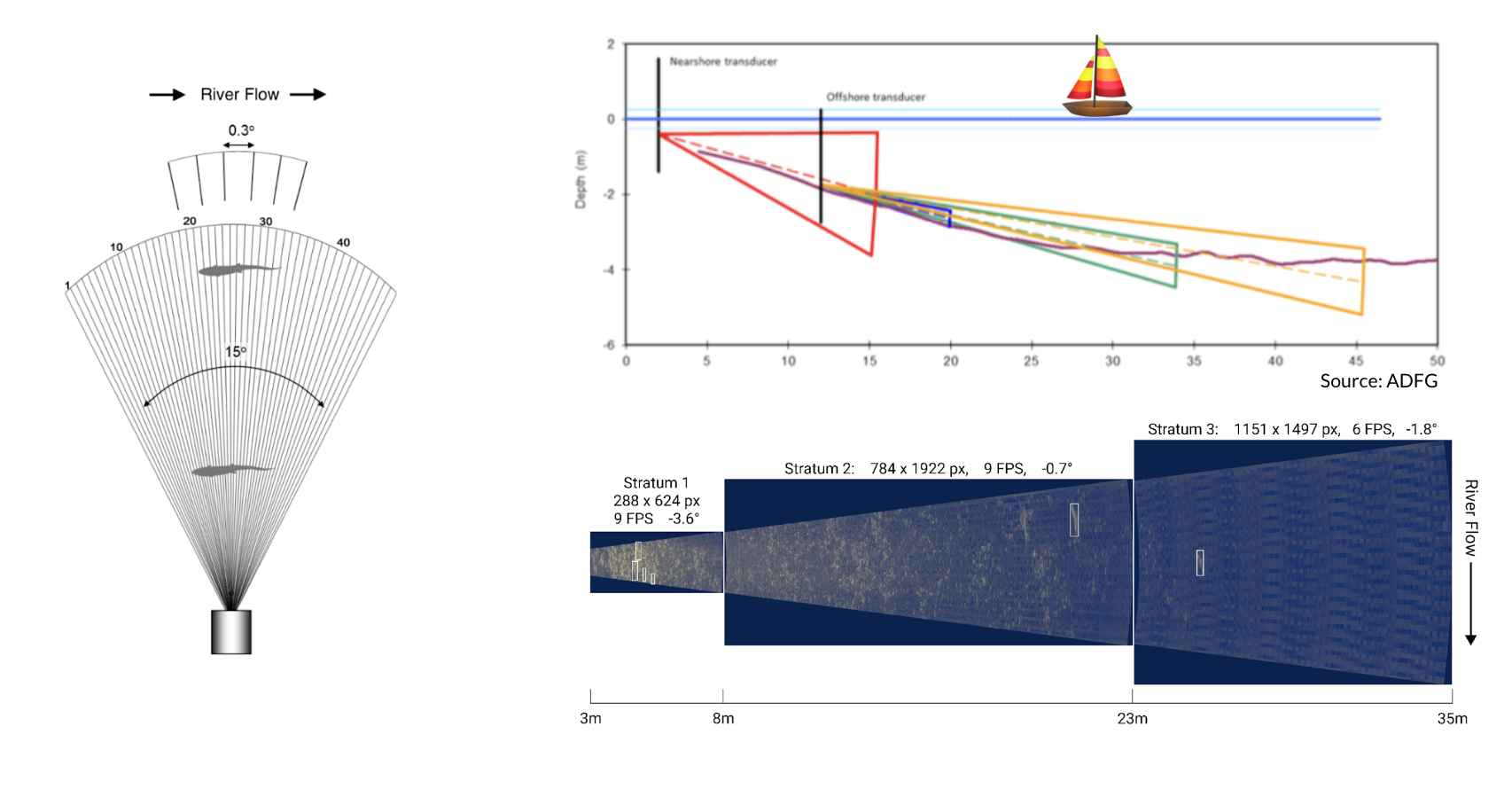}
	\centering
    \caption{Sonar camera arrangement: Here are some diagrams of the camera deployment. On the left, you can see the camera shooting out multiple acoustic sonar beams - they are used to pick up fish in high resolution. The closer the fish are to the camera, the higher their resolution is. In the top right, you can see how these sonar cameras are placed to "see" all areas where the salmon might swim. There are 2 sonar cameras - the one with the red triangle only captures one field of view; the one with the three narrow triangles oscillates between capturing three different stratas (20 min at one, 20 min at the second, 20 min at the third). The bottom right image shows what these three strata images look like when combined. The white boxes are the annotated fish swimming through the stream. Images have been provided from the Alaska Department of Fish and Game and Caltech\cite{cfc2022eccv}}
    \label{fig:caltech_fish_counting}
\end{figure}

\textbf{What is a use case?} Recent advances in hardware have allowed computer vision to expand to underwater locations. The Caltech Fish Counting task leverages sonar cameras placed in rivers to detect, track, and count salmon as they swim upstream \cite{cfc2022eccv}. The setup of these types of cameras within rivers is illustrated in \fig{fig:caltech_fish_counting}. They cannot rely on infrared sensors, so they capture images continuously across a specified period. Fisheries managers review the videos and manually count the number of salmon. Caltech researchers are working on automating this with computer vision \cite{cfc2022eccv}.

\textbf{What is the potential?} Computer vision has led to a set of technologies that can aid wildlife conservation across terrestrial, aquatic, and lab environments. Using computer vision as a tool can help solve limitations in manual data analysis by saving time and by limiting external bias. Processing large amounts of data quickly allows ecologists to then identify ecological patterns, trends, etc. in their scientific space and facilitates quicker lead times on field observations. Their science, then, informs ecological actions and goals. The integration of computer vision into wildlife conservation is dynamically automating animal ecology and conservation research using data-driven models. \cite{tuia_perspectives_2022}

\subsection*{Case Study 4: Game Theory and Optimization}

\noindent \textbf{Principle: Economics and Artificial Intelligence should be leveraged in conservation challenges to optimize decision-making}

Artificial intelligence is actively being used to combat wildlife threats. When designing conservation tools, like sensors, one key challenge is where to place them in an animal's ecosystem to collect relevant data. Researchers are looking into ways to leverage artificial intelligence methods to optimize conservation/resource planning and policy-making. One such field in computer science that differs from computer vision is the use of game theory for more effective data collection. \textbf{Game theory }is a collection of analytical tools that can be used to make optimal choices in inter-actional and decision-making problems. The use of game theory for conservation has only recently become a field of study. 

\textbf{What is its use?:} In non-mathematical terms, optimization is the study of how to make the best or most efficient decision given a certain set of constraints. In probability theory and machine learning, the multi-armed bandit problem is one type of optimization problem in which a limited set of resources must be split among/between competing choices to maximize expected gain. This problem is a sub-class of a broader set of problems called stochastic scheduling problems. In these problems, each machine provides a reward randomly from a probability distribution that is not known a-priori. The user's objective is to maximize the sum of the rewards. These techniques are commonly used for logistics (routing) coordination and financial portfolio design, though they have also been adapted to be used for modeling nefarious actors and optimally countering them. In wildlife scenarios, biologists often have to use a small number of tools to collect data in a vast environment often hundreds of square kilometers. The use of optimization strategies has recently begun to help ecologists and biologists pinpoint locations to effectively collect data descriptive of a large ecological habitat. 

\textbf{How is it used?}
\textit{Patrol Planning.} Wildlife poaching and trading threaten key species across ecosystems. Illegal wildlife trade facilitates the introduction of invasive species, land degradation, and biodiversity loss \cite{unep}. Historically, park rangers have recorded where poachers have struck. However, in most national parks, there is a limited supply of park rangers. They often are limited to driving, walking, or biking around the parks. Several parks have repositories of historical data detailing poaching locations identified in the past. This data can be used to predict likely poaching threats and locations in the future. Work has been done in the game theory and optimization space to leverage machine learning (on the historical data) and optimize multi-modal (i.e. driving and walking) patrol planning. Ultimately, parks and wildlife conservation organizations want to find the optimal answer to the questions, ``How should I organize my patrols?" and ``How will adversaries respond?" \cite{xu2021robust}. This optimization technique provides them with a way to answer those questions directly.

\textit{Economic Modeling.} Additional researchers, including Keskin and Nuwer, are working toward understanding the \textbf{economics} behind these wildlife threats. Poaching functions as an additional source of income for individuals in rural communities who may rely primarily on tourism for income. If these communities cannot rely on tourism, they may focus on wildlife trafficking, as those species are prevalent near them\cite{nuwer2018poached}. A review of wildlife tracking\cite{burcu2022wildlifesupplychain} focusing on operations and supply chain management recognized four challenges that limit preventative measures:
\begin{enumerate}
    \item the difficulties of understanding the true scale of illegal wildlife trade (IWT) from available data;
    \item the breadth of the issue - trafficked animals are used for food, status symbols, traditional medicine, exotic pets, and more (this requires the policy remedy to be multifaceted), and sometimes IWT operates in countries with corrupt governments or limited infrastructures for law enforcement and monitoring;
    \item IWT groups are geared toward undetectable operations, especially from financial institutions;
    \item IWT is considered less serious than other trafficking, i.e. human, drugs, weapons.
\end{enumerate}

There are several suggested ways to apply research in supply-chain operations toward combating IWT \cite{burcu2022wildlifesupplychain}. These include: bolstering data through satellite data, acoustic monitoring, news scraping, and finding online markets; strengthening data detection and prediction through network analysis and understanding data bias; modeling the problem as a network interdiction problem to see how to disrupt the supply chain network; more effective resource management and reducing corruption. By analyzing the complex supply chain and operations behind IWT, Keskin et al. illuminated a more clear picture of each location/scenario individually, which allows an informed and targeted response to prevent illegal wildlife trafficking\cite{krizhevsky2012ImageNet}.

\begin{figure}
    \centering
    \includegraphics[width=0.6\textwidth,page=1,trim=0.0in 0in 0in 0.0in,clip]{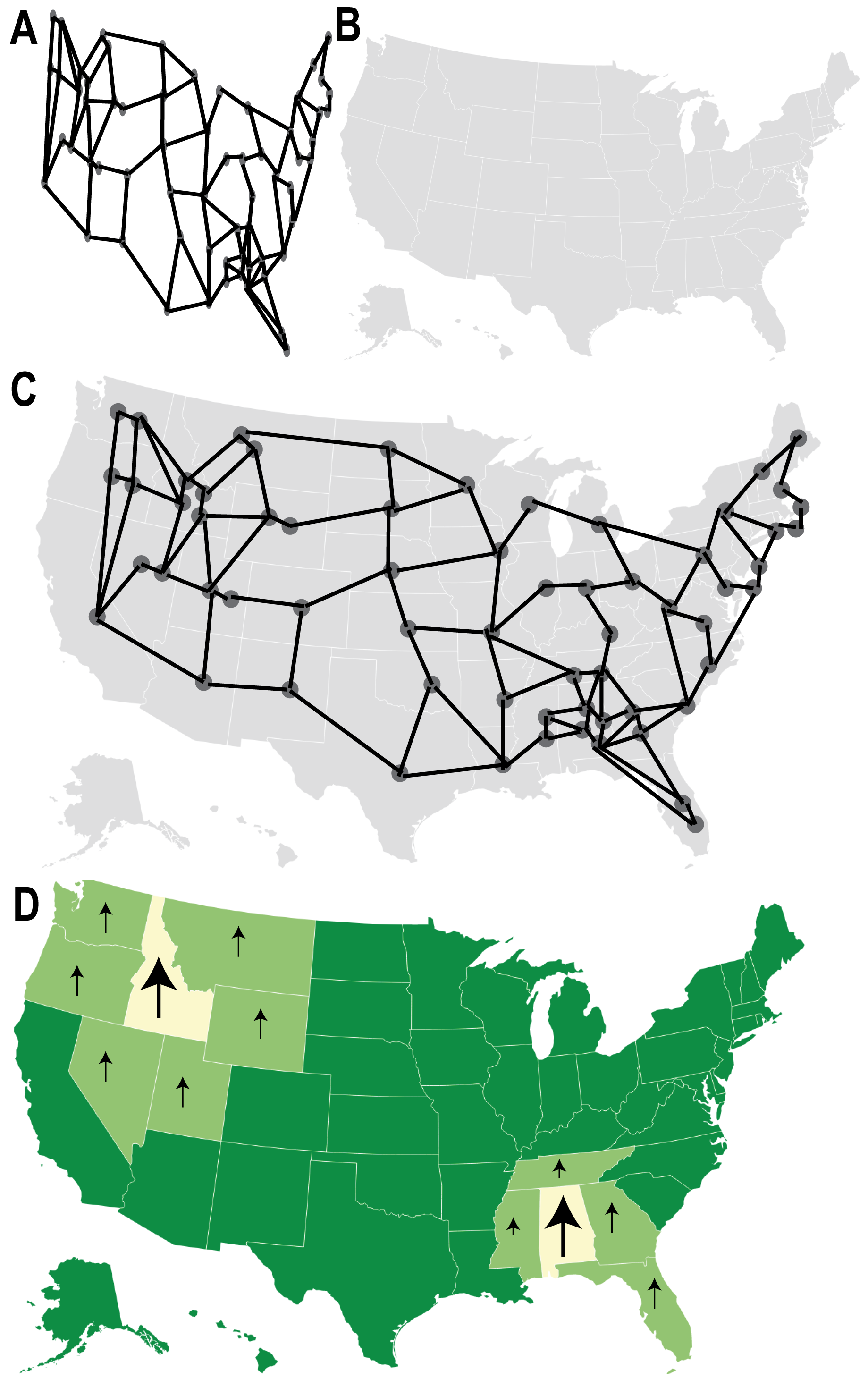}
    \caption{ Utilizing game theory and optimization for conservation practices.
    A) Data mapping of a conservation issue to determine which states conservation funding is most important. B) Raw map of the United States. C) Overlapped image of the clustering depicted in A with the raw map of the United States. D) Data interpreted map displaying large arrows in the states where the most conservation is needed with smaller arrows (in light green) displaying states where clustering is beginning. Images made using DataWrapper.}
    \label{fig:Optimization}
\end{figure}

\textbf{What are some use cases?} Evidence from parks in Uganda suggests that poachers are deterred by ranger patrols, illuminating the increased need for robust, sequential planning \cite{xu2021robust}. Computer science economists have worked on adversarial modeling to demonstrate poachers' deterrence to patrols along with other poacher behavior patterns \cite{xu2021robust}. An illustration of poaching patterns with increased patrols are shown in \fig{fig:Optimization}.

Researchers working at the Jilin Huangnihe National Nature Reserve in China first used machine learning to predict poaching threats and then used an algorithm to optimize a patrol route. When rangers were dispatched in December 2019, they successfully found forty-two snares, significantly more than they had found in previous months and patrols \cite{paws}. Combining machine learning and optimization techniques, therefore, has proven to increase the efficiency of patrol planning and can be expanded to more conservation management applications as well. 

\textbf{What is the potential?} Applying optimization techniques across conservation-oriented tasks will provide insight and better resource usage to historically under-resourced applications and programs. In addition, these optimization techniques and economically-focused viewpoints can prompt organizations and governments to identify and quell issues more efficiently. Programs can best utilize the limited resources they have and do so in an efficient data-driven manner. This can, in theory, be scaled to any resource-limited situation, too. Those with camera traps, for example, can study where to best place them to capture the most data-rich images. Those with limited AudioMoths, similarly, can study where to place them to ensure optimal and most realistic acoustic captures.

\subsection*{Case Study 5: Cat Collar}
\noindent \textbf{Principle: Solutions can be simple and should not be over-engineered}
This case study serves to provide an example that fails to fit the restrictive title of conservation technology that has taken hold and the overly technocentric vision that often results. A significant challenge to be considered when designing conservation tools is viability. A device or solution that is functional may not be enough. It must also be accessible in terms of parts availability and costs. The intended user base of any conservation tool is likely quite small. While the invention of a powerful tool may provide substantial functionality and/or opportunity, it is possible that (through traditional avenues) consumer demands and means are  insufficient to support its development and distribution. When a consumer cannot fully utilize or understand a conservation solution, it can fail, such as when there is little to no local adoption of the tools developed\cite{adugna_review_2021}. While frugal science is defined as reducing the cost of equipment in terms of bio-engineering, its driving principles are uniquely suited to conservation developments. Frugal science is subtly different than the Do-It-Yourself (DIY) and Free and Open-Source Hardware (FOSH) as it is focused on utilizing the most frugal ingredients to build your tool.  While not all devices can be designed frugally, many can have dramatically reduced costs as much of the conservation space is not purpose-built designs, but instead, like camera traps, are designed for hunters but also utilized also by ecologists and conservationists around the world. 

\textbf{What is its use?}  The collar is used to help combat nearly 1.3–4.0 billion birds per year killed by domestic cats in the US alone\cite{loss_impact_2013}. This makes domestic cats one of the nation's most significant anthropogenic threats to wildlife, yet very little counter-action has been taken. Common fondness towards cats makes enforcing restrictions on them difficult and makes enacting equivalent invasive species eradication methods extremely unpopular, even for feral cats\cite{loyd_public_2012}. Despite how silly the Birdsbesafe® collar looks, it has been found that collared cats killed 19 times fewer birds than did  their uncollared cats\cite{willson_birds_2015}. The collar is far less costly and controversial than alternative measures and allows cats to continue prowling freely. While no one would expect it at first glance, this demeaning accessory could be the most realistic, effective solution to preventing cat-caused bird extinction.

\textbf{How is it used?} The Birdsbesafe® collar is a simple woven fabric collar with bright collars painted along the fabric. The application of this device is simple as a large percent of the US population has outdoor cats. These collars are around $\$10$ and are made of two-inch wide cotton fabric tubes connecting to a quick-release collar. The collars are meant to be worn by both domesticated and wild cats as a means to reduce bird deaths. Much like other Felidae, domesticated cats (\textit{Felis catus}) have stealthy prey-stalking behavior. The collar has bright colors that can be seen from far away by both birds and small mammals that the cats may prey upon. This solution is simple as no technology is required, and the tool is used by hooking onto a collar around the cat's neck so the cat cannot pull them off. Additionally, these collars were designed with cats in mind as it does not impede on any of the cat's primary needs, including eating, drinking, sleeping, urinating, or defecating. Functionally this tool is a wearable technology that utilizes color to assist in the reduction of bird deaths, and it is as simple as hooking it around your pet cat's collar. 

\textbf{What is a use case?} In contrast to many of the previously discussed case studies, the idea of a conservation tool can be as simple as a brightly colored collar. In biomechanics, space footwear is designed to reduce the stress on the joints of the foot. This is verified using scientific data and techniques, but an essential factor in shoes is not just biomechanical support but how they look on your feet. This is the human element of human-centered design and is an important consideration when working with domestication species. When working with a domesticated species, like the cat, the solution amongst biologists is very simple: do not let your cat outside. However, the human element in this solution is that not all folks will follow these recommendations, and therefore other techniques need to be used. Less complex tools, like the cat collar, whose simplicity and low cost facilitate broader implementation, can make essential steps in conservation. This solution does not solve the issue of invasive cats killing birds and other animals but is dramatically reduces the impact that cats have on those that purchase these for their cats. 


\textbf{What is the potential?} Financing the development of conservation tools is a significant issue, and there may be few solutions that allow financing of crucial pilot tools and prototypes. Support from philanthropic organizations or technology organizations working in the technology space, such as Google Earth, Bezos Earth Fund, or AI4Good from Microsoft, can allow small startups to fund grants for these tool creations. But instead of this support, we encourage the implementation of frugal science methodology\cite{byagathvalli_frugal_2021}. 

\section*{Conclusions}
\begin{figure}
	\includegraphics[width=1\textwidth,page=1,trim=0.0in 0in 0in 0in,clip]{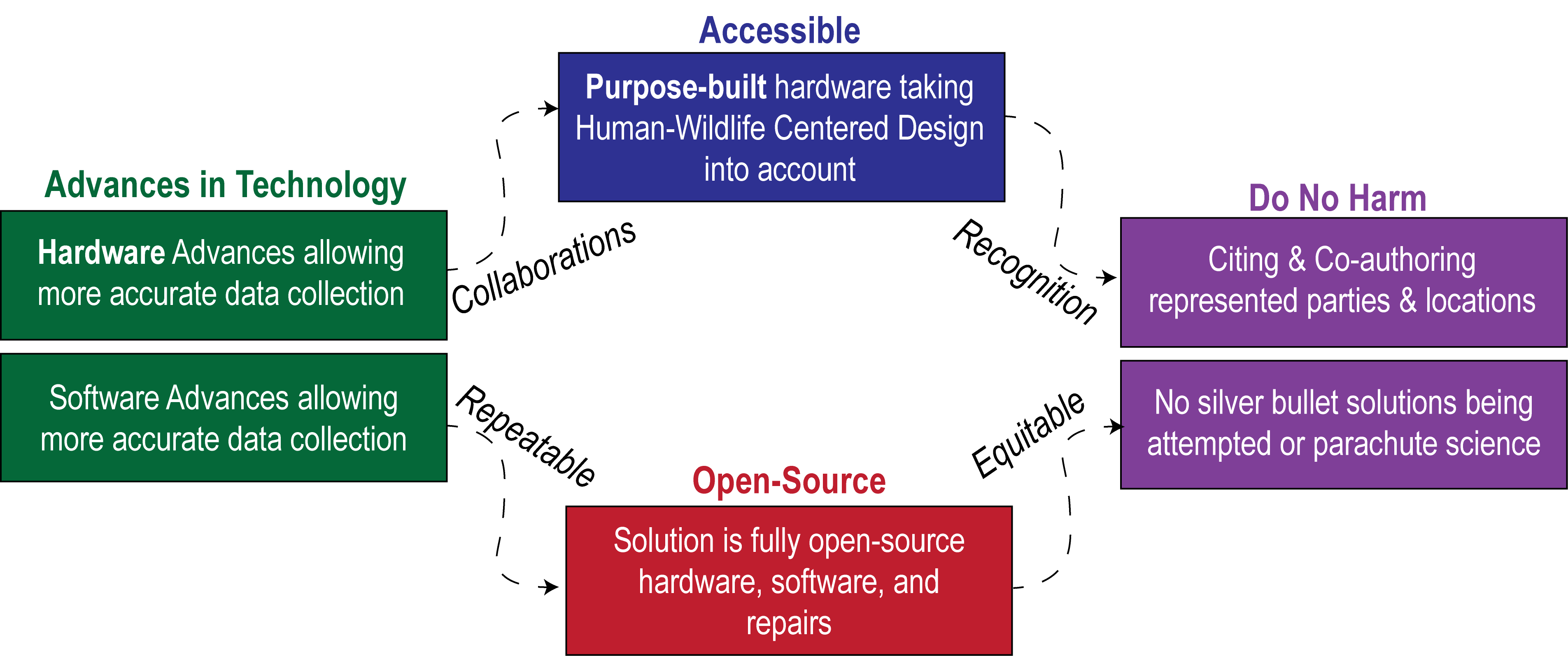}
	\centering
	\caption{Process that goes into designing and utilizing technology to develop new conservation tools. Silhouettes created by Gabriela Palomo-Munoz and Undraw.co.\\}
	\label{SummationFigure}
\end{figure}

\textbf{Conservation tools vary but are united in their potential to aid conservation.}
There is no single solution to the many challenges in conservation. Conservation tools are designed to be a part of a community's toolkit to help conserve and protect wildlife, and we discuss the key themes that make them successful in \fig{SummationFigure}. As these case studies show, conservation tools are not meant to solve all problems, but they can be useful in contexts where previous methods are too onerous or costly. Developers of conservation tools must understand that their designs need to be user-friendly for conservation practitioners and be viewed as a resource rather than a complete solution for addressing biodiversity decline. The most effective solutions are those that are realistically implementable and take into consideration the context of human-wildlife interactions in the design process. In this paper, we review five case studies of specific conservation tools that are advancing wildlife conservation. When examining these tools, it is important to consider the context in which they are used and the specific conservation issues they are addressing.


To develop effective conservation tools, biologists, computer scientists, and engineers must collaborate and apply their expertise. These interdisciplinary teams must also work with community members who have a deep understanding of conservation challenges. The wide range of perspectives and challenges addressed through these partnerships allow conservation tools to take many forms. We highlight five key characteristics of successful conservation tools. Open-source and accessible solutions like the AudioMoth offer opportunities for crowd-sourcing and additional improvements, as well as the ability to adapt existing frameworks to similar problems. Hardware from other fields can be repurposed in innovative ways to benefit conservation, such as using eDNA to reduce the invasiveness of data collection techniques. Existing software like computer vision can also be applied to the conservation field to streamline and expand data analyses. Successful conservation tools are not limited to biology, engineering, and computer science; they can also benefit from non-traditional fields like math for identifying ideal collection sites. Finally, not all solutions need to be high tech to be effective. Simple solutions, like cat collars with bells to protect birds, can also be effective conservation tools.


In this paper, we aim to provide a foundation for future conservation tool creators by reviewing case studies of successful tools and highlighting key themes. These case studies demonstrate the diverse range of approaches that can be taken in conservation technology, from simple cat collars to complex machine learning and game theory methodologies. By drawing on the expertise of interdisciplinary teams that include biologists, computer scientists, engineers, and community members, we can develop effective tools that address the unique challenges of each conservation context. As we work to conserve and protect wildlife, it is essential to remember that conservation tools are just one part of a larger toolkit and should be integrated into traditional and indigenous approaches to conservation. Through this review, we hope to inspire the development of innovative solutions to address the pressing needs of biodiversity conservation. Ultimately, conservation technology is essential for addressing the challenges of biodiversity preservation and promoting sustainable solutions for human-wildlife interactions.


\section*{Acknowledgements} 
Thank you to all of the members of the Georgia Tech Tech4Wildlife Student Organization for their support. 

\bibliographystyle{unsrt}

\clearpage

\newpage
\begin{table}[t]
		\centering
		\includegraphics[width=0.9\textwidth,page=1,trim=0.0in 0in 0in 0.0in,clip]{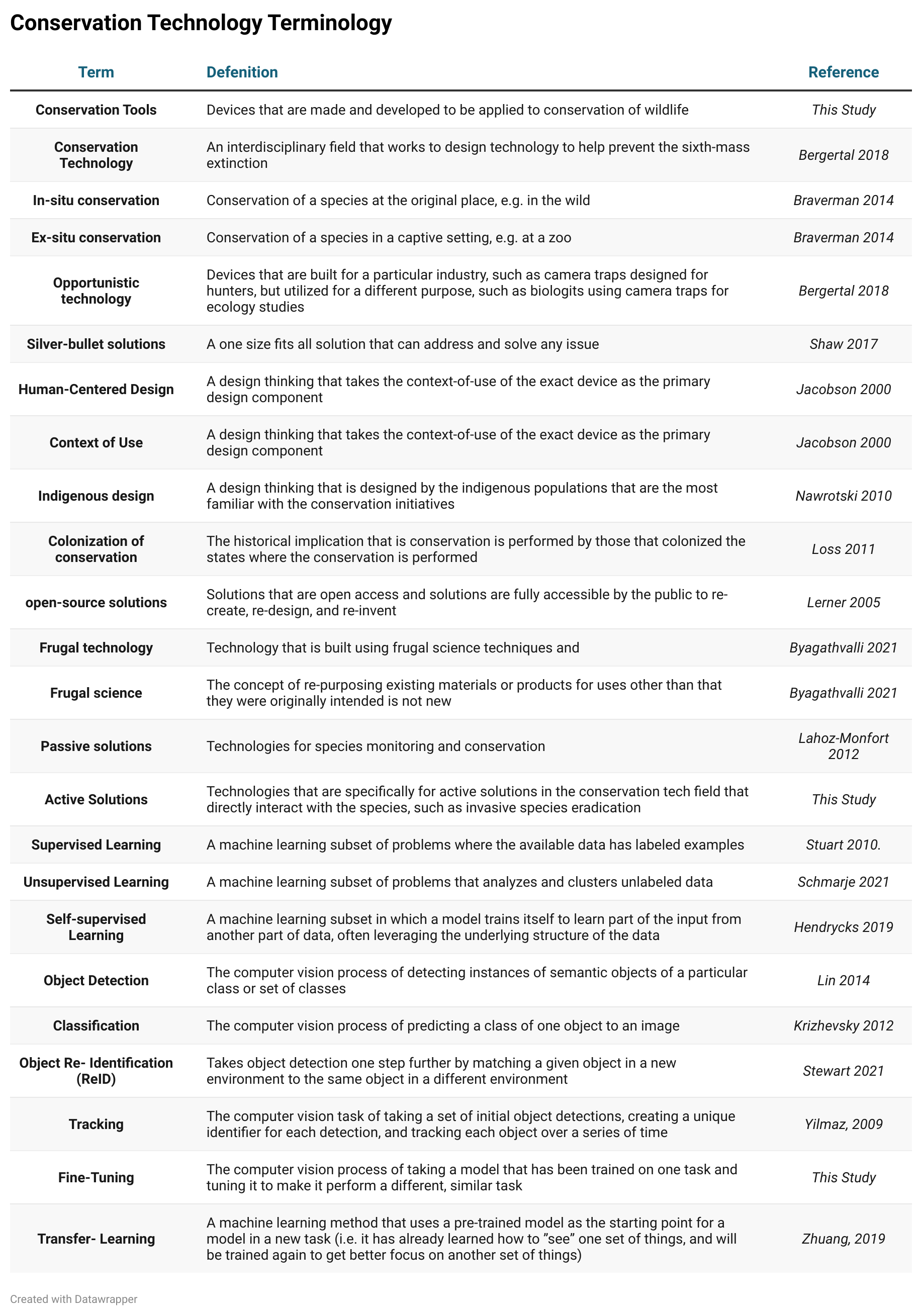}
		\label{suptab1}
\end{table}

\begin{table}[htb!]
		\centering
		\caption{Terms used by different groups practicing utilizing advanced and new technology to develop conservation tools and references to find more information about each term.
		}
		\label{tab:Table2_1}
\end{table}

\clearpage

\newpage

\end{document}